\documentclass[twocolumn,showpacs,superscriptaddress,amsmath,amssymb,prl]{revtex4}

\pdfoutput=1
\usepackage{dcolumn,graphicx,color,booktabs}

\renewcommand{\figurename}{Fig.}
\renewcommand{\tablename}{Table}
\makeatletter\renewcommand{\fnum@figure}[1]{\figurename~\thefigure.}\makeatother
\makeatletter\renewcommand{\fnum@table}[1]{\tablename~\thetable.}\makeatother
\usepackage[colorlinks,plainpages=false,linkcolor=black,urlcolor=blue,citecolor=blue,pdfpagemode=UseNone,pdfstartview=FitH]{hyperref}

\begin{document}

\title{Chiral honeycomb superstructure, parquet nesting, and Dirac cone formation in Cu-intercalated 2H-TaSe$_2$}

\author{A.~A.~Kordyuk}
\affiliation{Institute for Solid State Research, IFW-Dresden, P.O.Box 270116, D-01171 Dresden, Germany}
\affiliation{Institute of Metal Physics of National Academy of Sciences of Ukraine, 03142 Kyiv,  Ukraine}

\author{D.~V.~Evtushinsky}
\author{V.~B.~Zabolotnyy}
\author{T.~Haenke}
\author{C.~Hess}
\author{B.~B\"{u}chner}
\affiliation{Institute for Solid State Research, IFW-Dresden, P.O.Box 270116, D-01171 Dresden, Germany}

\author{A.~N.~Yaresko}
\affiliation{Max-Planck-Institute for Solid State Research, 70569 Stuttgart, Germany}

\author{H.~Berger}
\affiliation{Institut de Physique de la Matiere Complexe, EPFL, 1015 Lausanne, Switzerland}

\author{S.~V.~Borisenko}
\affiliation{Institute for Solid State Research, IFW-Dresden, P.O.Box 270116, D-01171 Dresden, Germany}

\begin{abstract}
Powered by effective parquet nesting a commensurate chiral honeycomb superstructure in a trigonally packed transition metal dichalcogenide TaSe$_2$ results in a Dirac cone anomaly in one-particle excitation spectrum. However, the formation of the well defined Dirac point seems to be hindered by effective scattering on 2D plasmons.
\end{abstract}


\maketitle

Dimensional transitions in electronic systems is in one of the main focuses of modern condensed matter physics \cite{VallaN2002, FengN2008}. The renowned example is the transformation of the quasi two-dimensional (2D) electronic structure of graphite into the strictly 2D structure of graphene, where lowering the dimensionality turns the semi-metal with a small band gap into a zero gap semiconductor with a Dirac cone anomaly \cite{NovoselovN2005, KatsnelsonNP2006, NovoselovNM2007}. This has brought into play a number of new physical phenomena and has already created a new direction in a high-end technology named "bandgap engineering" \cite{NovoselovNM2007}. The superstructural transitions provide another tool for band engineering \cite{BorisenkoPRL2008}. In this paper we show that intercalating 2H-TaSe$_2$ crystal \cite{WilsonAiP1975} with copper causes formation of a chiral honeycomb superstructure which, in turn, creates a Dirac cone anomaly in one-particle excitation spectrum, similar to the one discussed in case of graphene \cite{BostwickNP2007} or in topological insulators \cite{HsiehN2008}. The spectra analysis of the anomaly reveals a dramatic increase of the quasiparticle scattering when approaching the Dirac point that allows us to conclude in favour of plasmon scattering \cite{BostwickNP2007} that hinders the formation of the well defined Dirac point in 2D metals. The observed superstructure itself deserves a special attention: it presents an example of extremely stable chiral electronic ordering on the surface powered by an effective parquet nesting mechanism and can be interesting for various technical applications \cite{RavalCSR2009}.

In contrast to 1T polytype transition metal dichalcogenides, which are often called semi-metals and also known as excitonic insulators \cite{CercellierPRL2007}, the 2H polytypes are good metals with the large Fermi surface (FS) formed by two electronic bands crossing the Fermi level \cite{BorisenkoPRL2008, BorisenkoPRL2009}. The crystal structure of both polytypes is rather simple. It consists of three atom layer sandwiches: two layers of chalcogen atoms (Se, S, Te) and a layer of metal atoms (Ta, Nb, Ti) in between, all hexagonally packed. Different is the coordination of the metal atoms: octahedral for 1T- and trigonal prismatic for 2H-polytypes, respectively \cite{WilsonAiP1975}. Figure 1a shows large FS of 2H-TaSe$_2$ measured by angle resolved photoemission (ARPES) (see also Refs. \onlinecite{BorisenkoPRL2008, EvtushinskyPRL2008, InosovPRB2009}). One band forms the FS barrels (hole pockets) around the center and the corners of the Brillouin zone (BZ) while the other band forms the "dog bone" electron pockets. These bands are originated from the Ta 5d electrons and are doubled due to 2H coordination. Such a band structure has a propensity to fold causing additional electronic orderings (such as commensurate and incommensurate CDW in 2H-TaSe$_2$) which have been found to be extremely sensitive to the very details of the FS shape \cite{BorisenkoPRL2008}, and, consequently, can be tuned by intercalation \cite{WilsonAiP1975}, doping \cite{MorosanNP2006}, pressure \cite{SiposNM2008} or temperature \cite{InosovPRB2009}.

\begin{figure*}
\begin{center}
\includegraphics[width=0.8\textwidth]{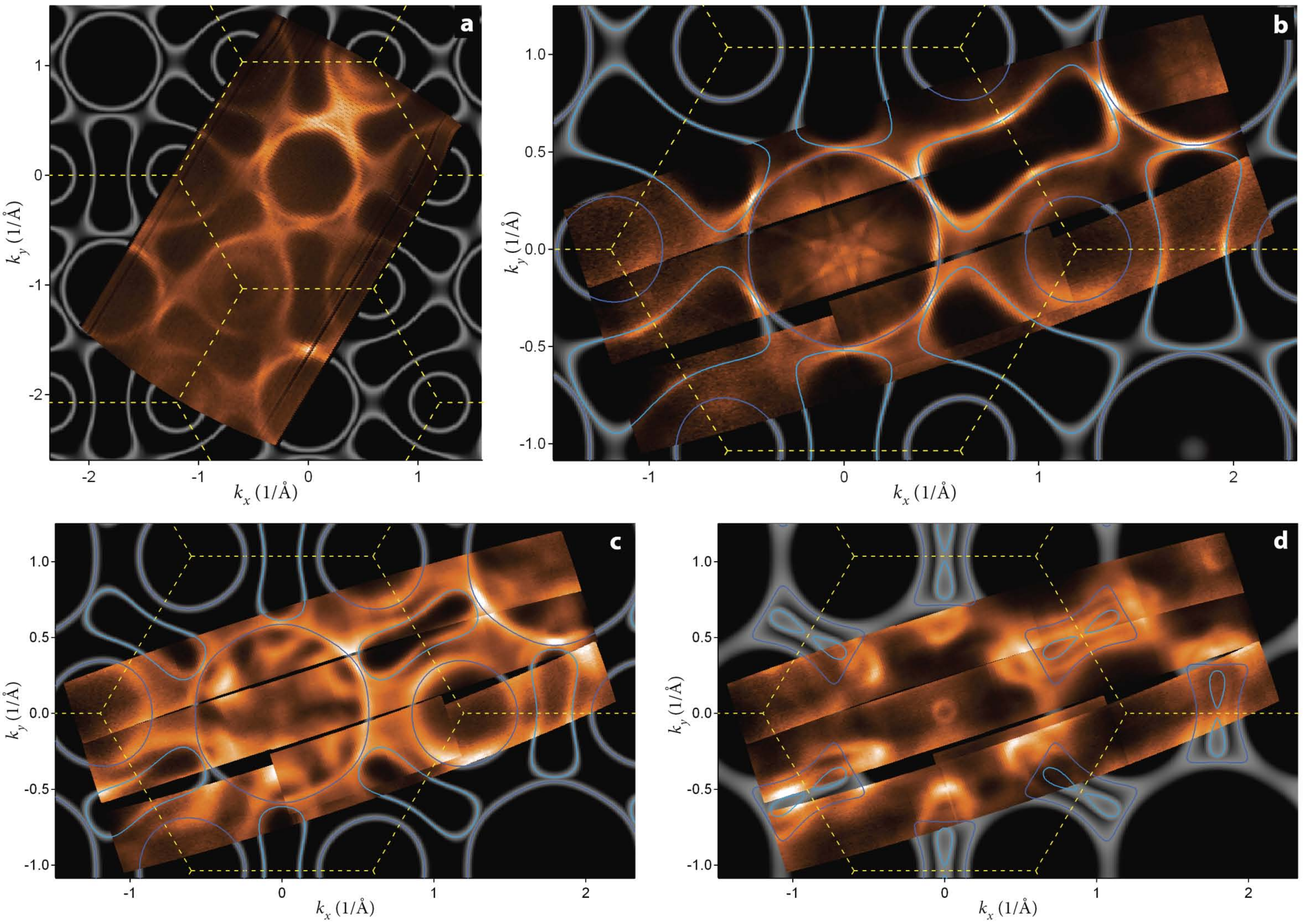}\\
\caption{\label{Fig1} Effect of structural modulation on Fermi surface (FS): Intensity maps of parent 2H-TaSe$_2$ (a) and Cu-intercalated 2H-TaSe$_2$ (b-d).  Experimental data are represented in gold color scale; the band fitting result is shown in gray color scale and by blue contours. Yellow dashed lines delimit the Brillouin zones (BZs). At the Fermi level (b), the new order is only seen as a chiral star-like feature. At higher binding energies, 0.16 eV (c) and 0.32 eV (d), the folding is seen over the whole BZ, while in the centre of the BZ only a single cone anomaly is present.}
\end{center}
\end{figure*}

Figure 1 b shows the FS of 2H-TaSe$_2$ intercalated by Cu. At first glance, the only principal difference with the FS of the parent compound is a star-like feature in the center of the Brillouin zone that is a sign of a new order. One may easily notice a remarkable peculiarity of the star: it is slightly rotated with respect to the main FS (its symmetry planes do not coincide with the symmetry planes of the crystal). In other words, the new order breaks the mirror symmetry of the crystal and introduces chirality. The chirality implies that the left and right ordered domains can exist but the fact that the chiral structure is observed in the reciprocal space indicates that the domain of one sign dominates in the real space.

\begin{figure}[t]
\begin{center}
\includegraphics[width=0.47\textwidth]{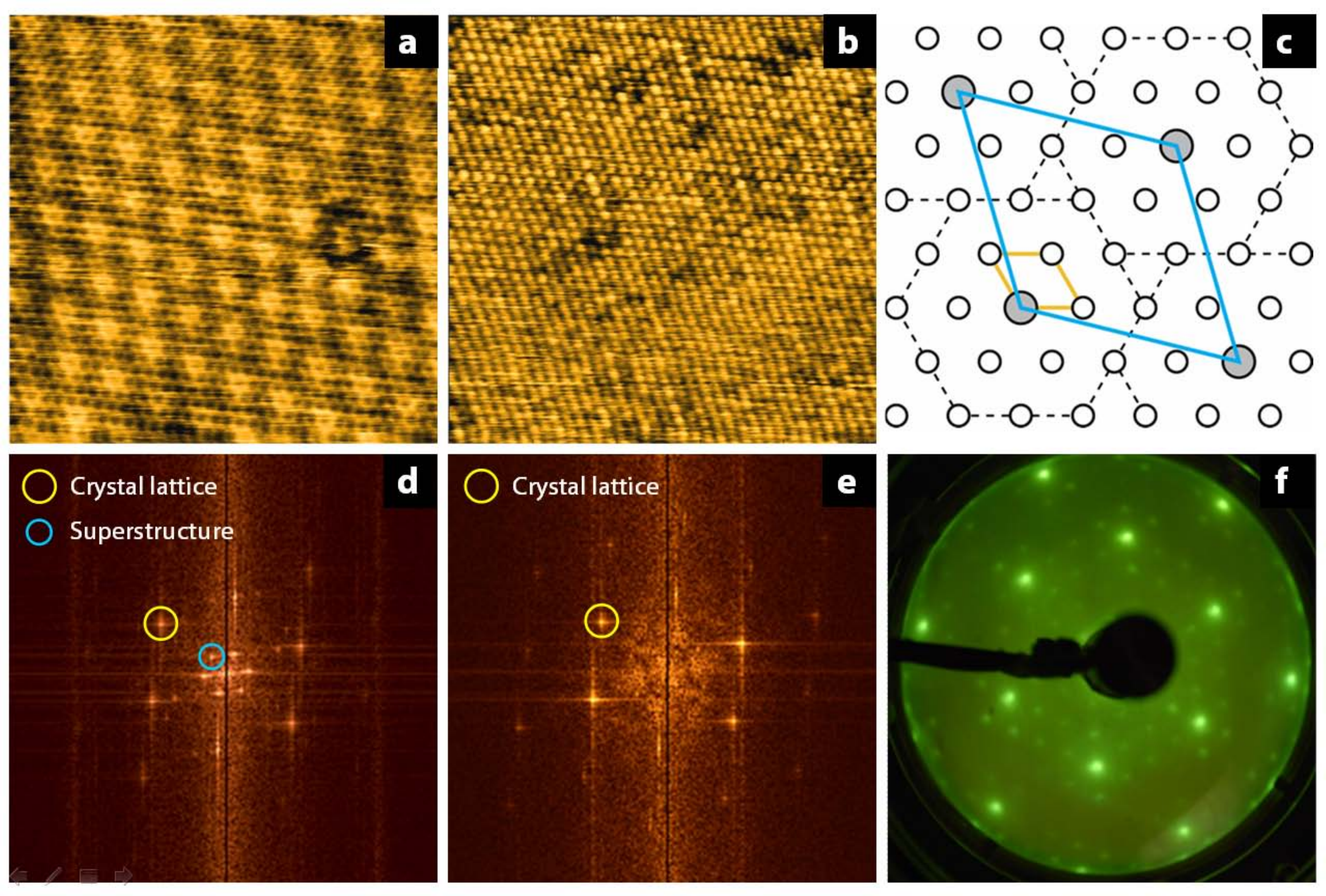}\\
\caption{\label{Fig2} Chiral honeycomb superctructure in real and reciprocal space. The real space images (a) and (b) are recorded by the scanning tunneling microscope (STM) in the field of view 10x10 nm$^2$ with +5 mV and +500 mV bias voltage, respectively. The reciprocal space images are obtained by Fourier transformed STM (d, e) and by low energy electron diffraction (LEED) (f). All data are recorded at room temperature. The chiral superstucture is observed over the entire low bias voltage range of $\pm$400 mV but disappears at higher bias. The super-cell is delimited by blue line on panel (c): it contains 13 Ta atoms and is rotated clockwise by about 13$^{\circ}$54' with respect to the original unit cell (yellow).}
\end{center}
\end{figure}

The symmetry of this order is clearly revealed by the scanning tunneling microscopy (STM) and low energy electron diffraction (LEED) measurements as shown in Fig. 2. The new unit cell contains 13 Ta atoms and is rotated clockwise by about 13$^{\circ}$54' with respect to the original unit cell (Fig. 1c). The new order can be described precisely as [(3 $-1$) (1 4)] superstructure, in matrix notation, or as ($\sqrt{13}\times\sqrt{13}$)R13$^{\circ}$54' superstructure, in Wood's notation, and is known as "$\sqrt{13}$ superstructure" or "star of David superstucture" that had been observed earlier for 1T-polytypes \cite{WilsonAiP1975}. The band folding simulation (Fig. 3a), which is based on hopping parameters derived from the experimental dispersions measured by ARPES, proves that the star-like feature in the center of the BZ is, in fact, constructed by the first replicas of the non-disturbed electronic structure.

Since the chiral superstructure of only one sign is realized on the surface of the whole single crystal (about 3$\times$3 mm) and is observed in the whole measured temperature range (from 15 to 300 K) one can conclude that the new order is very favorable energetically. This makes it very attractive from the practical point of view \cite{RavalCSR2009}, while at the fundamental level, the $\sqrt{13}$ superlattice in Cu-intercalated 2H-TaSe$_2$ appears puzzling in both its origin and its consequences.

To understand the origin of a superstructure one should answer two questions: where does the superstructure gain the energy from and what is the microscopic structure of the superlattice. Exploring the distribution of the electronic density below the Fermi level, we see where the gain of the kinetic energy of the electrons comes from: the folded electronic structure is effectively gaped. The maps of the electronic density at 0.16 eV and 0.32 eV binding energies (Fig. 1c,d) show that the replicas of the band structure are gapped everywhere they are observed except of the BZ centre.

\begin{figure}[t]
\begin{center}
\includegraphics[width=0.5\textwidth]{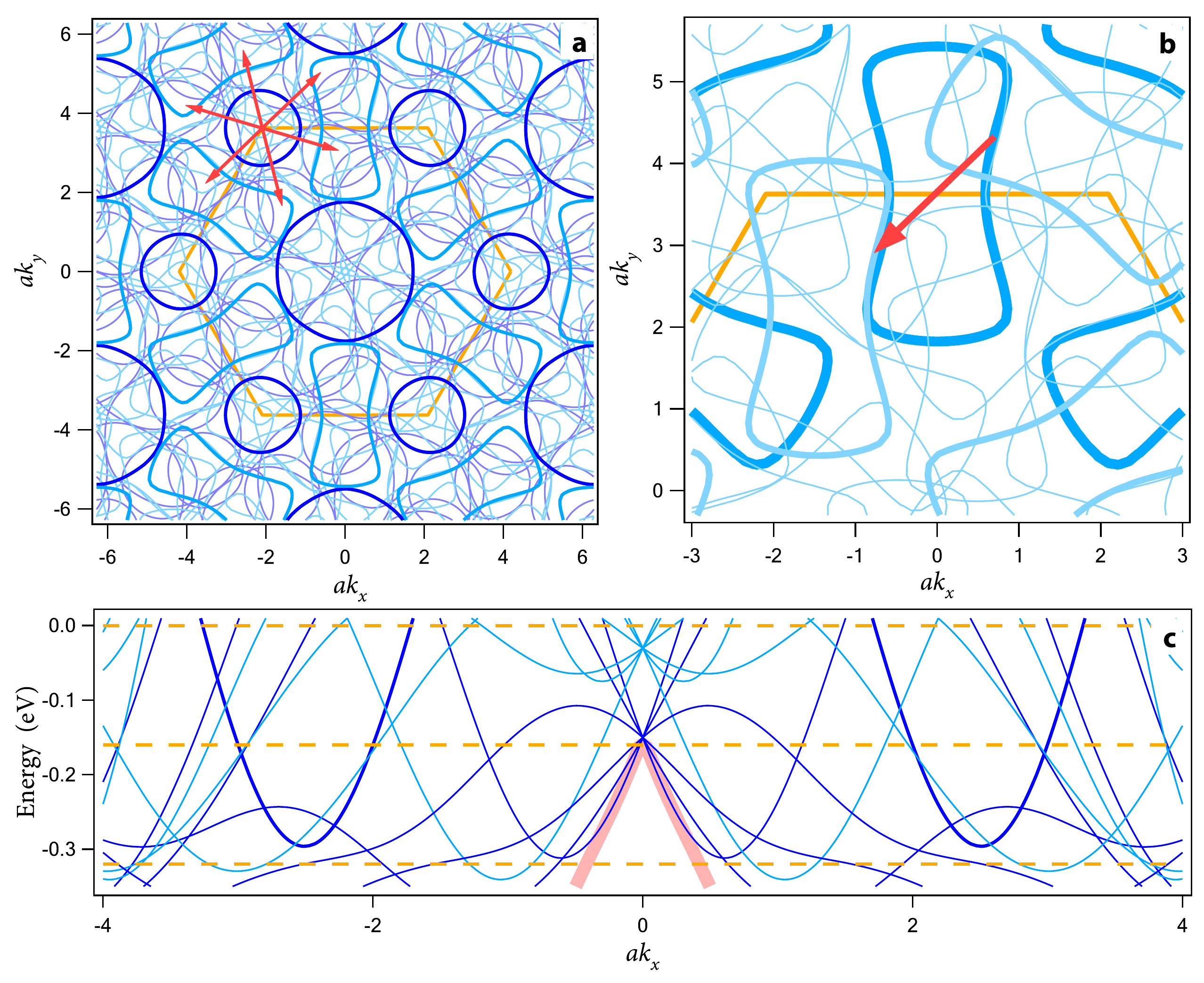}\\
\caption{\label{Fig3} Band folding and parquet nesting. Six-fold replication of the non-disturbed electronic band structure by $\sqrt{13}$R13$^{\circ}$54' vectors (red arrows) (a) proves that the star-like feature in the center of the BZ is, in fact, constructed by the first replicas. A precise parquet nesting of the "dog bone" FS sheets by exactly $\sqrt{13}$R13$^{\circ}$54' vectors (b) leads to effective gap opening over the large area in the reciprocal space. (d) The momentum-energy cut of the electronic band structure at $k_y = 0$ shows that the Dirac cone anomaly (fat red line), which is further explored in Fig. 4, is formed in the center of the BZ by first six replicas of the bonding band like an iris diaphragm in a lens is formed by blades. In all panels, bold contours correspond to original electronic bands and thin contours correspond to their first replicas (in panel b one of the replicas is made thicker for clarity); the bands which form the FS barrels and "dog bones" are shown by dark and light shadows of blue, respectively.}
\end{center}
\end{figure}

\begin{figure*}[t]
\begin{center}
\includegraphics[width=0.8\textwidth]{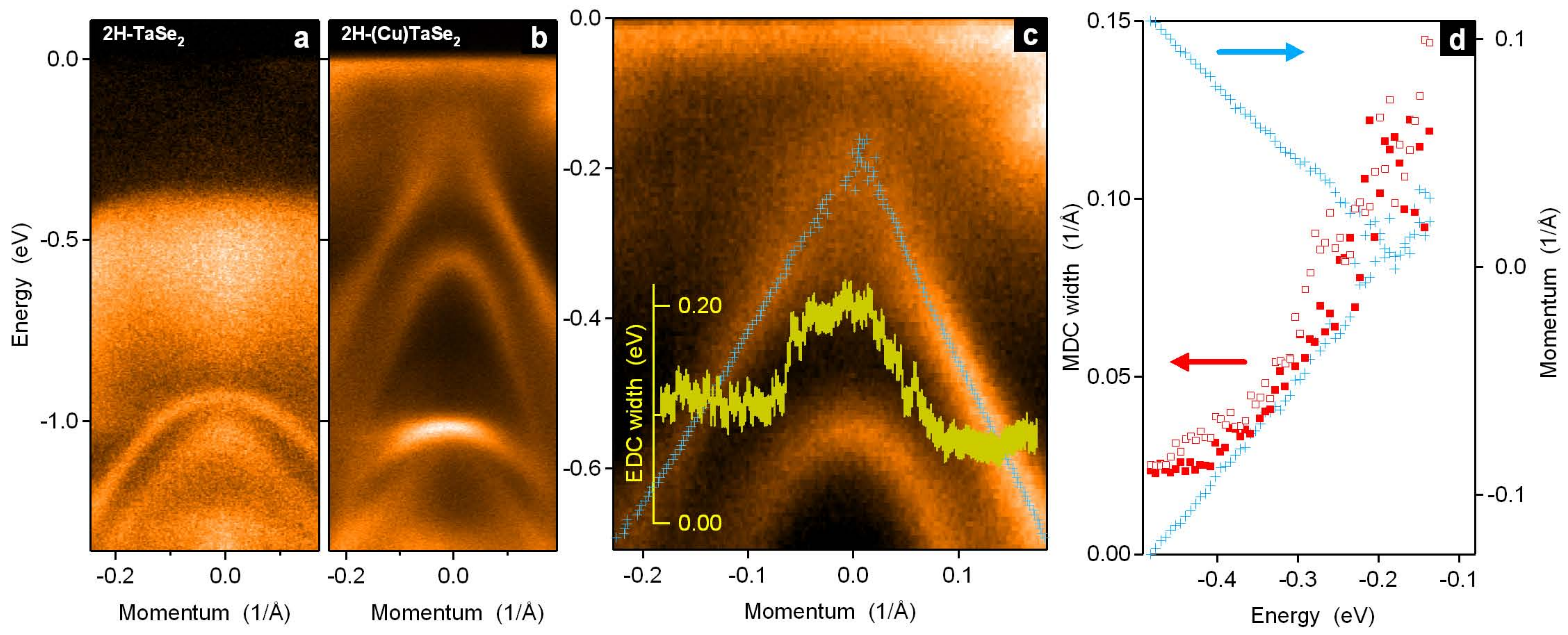}\\
\caption{\label{Fig4} Dirac cone formation and "evaporation". The momentum-energy cut through the center of the BZ of Cu-intercalated 2H-TaSe2 shows that below the Fermi level, the electronic excitations form a cone (b), a completely new feature comparing to the parent 2H-TaSe2$_2$ (a). The dispersion of the excitations in question, as follows from the positions of the momentum distribution curve (MDC) maxima shown by blue crosses on a zoomed image of the cone (c), is, in fact, linear. The width of both energy distribution curves (EDC) (c) and MDC (d), which are proportional to the scattering rate, tend to diverge approaching the cone vertex-the Dirac point.}
\end{center}
\end{figure*}

Figure 3b shows the reason for such a gap: a precise parquet nesting of the dog-bone FS sheet by exactly the $\sqrt{13}$R13$^{\circ}$54' vector. The segments of the FS are nested when, being shifted by a nesting vector, they coincide having the opposite Fermi velocities. The parquet nesting is the nesting of the segments within one FS sheet. It is efficient when the FS has extended segments of similar shape but opposite curvature. Naturally, such a good parquet nesting should result in a larger gap and higher transition temperature than the inter-sheet FS nesting responsible for 3$\times$3 CDW in 2H-TaSe$_2$ (see Ref. \onlinecite{BorisenkoPRL2008}). The gap value can be estimated as 90 meV while the transition temperature remains beyond 300 K, the maximal temperature reached in our experiment.

Revealing the microscopic picture of the superlattice is more intricate. The observed gain in the electronic kinetic energy can cause the superstructural transition of the crystal lattice [the structural transition which leads to the formation of periodic lattice distortions \cite{FriendJPC1979} (PLD)] or non-PLD ordering, such as ordering of the intercalate \cite{WilsonAiP1975} or a self-sufficient electronic density wave. In addition, the superlattice can be formed in the bulk of the crystal and/or on its surface.

Comparing the intensity maps shown in Fig. 1a with the simulations of the folding one can conclude that the former represent the superposition of the electronic densities from two types of crystal lattice: with and without the superstructure. Since, at the excitation energy used in our experiment, ARPES accesses the electrons from more than one layer \cite{ZabolotnyyPRB2007}, one can further conclude that the superstructure is realized either in the bulk or at the surface but not in both (the domain scenario is not applicable here due to the fact that we routinely observe the chiral superstructure of only one sign). Clear observation of the superstructure by surface sensitive STM shows that it is the surface layer that is reconstructed. This rules out the ordering of intercalate. In addition, the superstructure, as seen by STM, disappears at higher bias voltage, in contrast to $\sqrt{13}$ superstructure in 1T-TaSe$_2$ \cite{StoltzPRB2007}, which makes the surface PLD very unlikely. Therefore, we conclude that the microscopic origin for the superlattice is surface electronic density modulation realized on low binding energy electrons.

Now we switch to an interesting consequence of the $\sqrt{13}$ superstructure in Cu-intercalated 2H-TaSe$_2$---a Dirac cone formation. The bright point in the center of the BZ in Figure 1c appears to be a vertex of a cone in the electron excitation dispersion. The evolution of this point into a circle at higher binding energy is seen in Figure 1d. From the momentum-energy cut, taken around BZ center along $k_y = 0$ direction and examined in Figure 4, one may conclude that the dispersion is really linear, which is a signature of the Dirac fermions \cite{NovoselovN2005, KatsnelsonNP2006}. As a matter of fact, the cone is constructed by the six first-order replicas of the band which forms the FS barrels around the center and corners of the BZ (see Fig. 3c). It remains puzzling why the intensity of the electrons that forms the Dirac cone is dominating as compared to the intensity of the other replicas that should be present nearby in the momentum space. Intuitively, the formation of Dirac cones in electronic dispersion may be a general consequence of the honeycomb symmetry \cite{ShimaPRL1993} that, in case of graphene, is formulated in terms of symmetry between two equivalent carbon sublattices \cite{NovoselovN2005}. We believe that accurate analysis of the scattering potential \cite{ZabolotnyyEPL2009} will clarify this issue.

While a single surface state Dirac cone is considered as a signature of topological insulator \cite{HsiehN2008}, the further analysis suggests that well defined Dirac point is not allowed in a good metal such as 2H-TaSe$_2$. Figure 4d shows that the scattering rate, which can be deduced from the width of the ARPES spectrum \cite{KordyukPRB2005}, tends to diverge approaching the Dirac point, which looks as "evaporation" of the vertex of the cone in Fig. 1b,c. This observation supports the idea of strong electron-plasmon coupling suggested to explain a "kinked" dispersion in graphene \cite{BostwickNP2007}. Strong electron-plasmon coupling is indeed expected for any 2D metals \cite{PashitskiiLTP2008}. The coupling between Dirac fermions and 2D plasmons is expected to be even larger because the plasmons with the square-root dispersion scatter effectively the electronic excitations around the Dirac point \cite{BostwickNP2007}. Therefore, it is natural that the effect of plasmons on Dirac fermions observed in graphene is dramatically increased in such a good metal as 2H-TaSe$_2$.

In summary, we explore the mechanism and the consequences of the chiral honeycomb superstructure in 2H-TaSe$_2$ induced by Cu intercalation. Based on the observed spectral weight redistribution, we conclude that the new electronic ordering appears at the crystal surface due to effective parquet nesting. An interesting result of the new order is the formation of a Dirac cone anomaly, though with not well defined Dirac point.

The project is part of the Forschergruppe FOR538, and also supported by DFG BO 1912/2-1, the Swiss National Foundation for Scientific Research and by the NCCR MaNEP. We acknowledge discussions with Alexander Gabovich, Alex Gruneis, Dmytro Inosov, Changyoung Kim, Timur Kim, Martin Knupfer, Yuri Kopaev, Eugene Krasovskii, Igor Mazin, Ernst Pashitskii, Roman Schuster, and Vladimir Strokov.


\begin{thebibliography}{10}

\bibitem{VallaN2002}
{ T.~Valla, P.~D. Johnson, Z.~Yusof, B.~Wells, \textit{et~al}.,
  \href{http://dx.doi.org/10.1038/nature00774}{
\newblock Nature {\bf 417}, 627 (2002)}.}

\bibitem{FengN2008}
{ J.~Feng, R.~G. Hennig, N.~W. Ashcroft, and R.~Hoffmann,
  \href{http://dx.doi.org/10.1038/nature06442}{
\newblock Nature {\bf 451}, 445 (2008)}.}

\bibitem{NovoselovN2005}
{ K.~S. Novoselov, A.~K. Geim, S.~V. Morozov, D.~Jiang, \textit{et~al}.,
  \href{http://dx.doi.org/10.1038/nature04233}{
\newblock Nature {\bf 438}, 197 (2005)}.}

\bibitem{KatsnelsonNP2006}
{ M.~I. Katsnelson, K.~S. Novoselov, and A.~K. Geim,
  \href{http://dx.doi.org/10.1038/nphys384}{
\newblock Nat. Phys. {\bf 2}, 620 (2006)}.}

\bibitem{NovoselovNM2007}
{ K.~Novoselov, \href{http://dx.doi.org/10.1038/nmat2006}{
\newblock Nat. Mater. {\bf 6}, 720 (2007)}.}

\bibitem{BorisenkoPRL2008}
{ S.~V. Borisenko, A.~A. Kordyuk, A.~N. Yaresko, V.~B. Zabolotnyy,
  \textit{et~al}., \href{http://link.aps.org/abstract/PRL/v100/e196402}{
\newblock Phys. Rev. Lett. {\bf 100}, 196402 (2008)}.}

\bibitem{WilsonAiP1975}
{ J.~A. Wilson, F.~D. Salvo, and S.~Mahajan,
  \href{http://www.informaworld.com/10.1080/00018737500101391}{
\newblock Advances in Physics {\bf 24}, 117 (1975)}.}

\bibitem{BostwickNP2007}
{ A.~Bostwick, T.~Ohta, T.~Seyller, K.~Horn, and E.~Rotenberg,
  \href{http://dx.doi.org/10.1038/nphys477}{
\newblock Nat. Phys. {\bf 3}, 36 (2007)}.}

\bibitem{HsiehN2008}
{ D.~Hsieh, D.~Qian, L.~Wray, Y.~Xia, \textit{et~al}.,
  \href{http://dx.doi.org/10.1038/nature06843}{
\newblock Nature {\bf 452}, 970 (2008)}.}

\bibitem{RavalCSR2009}
{ R.~Raval, \href{http://dx.doi.org/10.1039/b800411k}{
\newblock Chem. Soc. Rev. {\bf 38}, 707 (2009)}.}

\bibitem{CercellierPRL2007}
{ H.~Cercellier, C.~Monney, F.~Clerc, C.~Battaglia, \textit{et~al}.,
  \href{http://link.aps.org/abstract/PRL/v99/e146403}{
\newblock Phys. Rev. Lett. {\bf 99}, 146403 (2007)}.}

\bibitem{BorisenkoPRL2009}
{ S.~V. Borisenko, A.~A. Kordyuk, V.~B. Zabolotnyy, D.~S. Inosov,
  \textit{et~al}., \href{http://link.aps.org/abstract/PRL/v102/e166402}{
\newblock Phys. Rev. Lett. {\bf 102}, 166402 (2009)}.}

\bibitem{EvtushinskyPRL2008}
{ D.~V. Evtushinsky, A.~A. Kordyuk, V.~B. Zabolotnyy, D.~S. Inosov,
  \textit{et~al}., \href{http://link.aps.org/abstract/PRL/v100/e236402}{
\newblock Phys. Rev. Lett. {\bf 100}, 236402 (2008)}.}

\bibitem{InosovPRB2009}
{ D.~S. Inosov, D.~V. Evtushinsky, V.~B. Zabolotnyy, A.~A. Kordyuk,
  \textit{et~al}., \href{http://link.aps.org/doi/10.1103/PhysRevB.79.125112}{
\newblock Phys. Rev. B {\bf 79}, 125112 (2009)}.}

\bibitem{MorosanNP2006}
{ E.~Morosan, H.~W. Zandbergen, B.~S. Dennis, J.~W.~G. Bos, \textit{et~al}.,
  \href{http://dx.doi.org/10.1038/nphys360}{
\newblock Nat. Phys. {\bf 2}, 544 (2006)}.}

\bibitem{SiposNM2008}
{ B.~Sipos, A.~F. Kusmartseva, A.~Akrap, H.~Berger, \textit{et~al}.,
  \href{http://dx.doi.org/10.1038/nmat2318}{
\newblock Nat. Mater. {\bf 7}, 960 (2008)}.}

\bibitem{FriendJPC1979}
{ R.~H. Friend and D.~Jerome, \href{http://stacks.iop.org/0022-3719/12/1441}{
\newblock Journal of Physics C: Solid State Physics {\bf 12}, 1441 (1979)}.}

\bibitem{ZabolotnyyPRB2007}
{ V.~B. Zabolotnyy, S.~V. Borisenko, A.~A. Kordyuk, D.~S. Inosov,
  \textit{et~al}., \href{http://link.aps.org/abstract/PRB/v76/e024502}{
\newblock Phys. Rev. B {\bf 76}, 024502 (2007)}.}

\bibitem{StoltzPRB2007}
{ D.~Stoltz, M.~Bielmann, M.~Bovet, L.~Schlapbach, and H.~Berger,
  \href{http://link.aps.org/abstract/PRB/v76/e073410}{
\newblock Phys. Rev. B {\bf 76}, 073410 (2007)}.}

\bibitem{ShimaPRL1993}
{ N.~Shima and H.~Aoki, \href{http://link.aps.org/abstract/PRL/v71/p4389}{
\newblock Phys. Rev. Lett. {\bf 71}, 4389 (1993)}.}

\bibitem{ZabolotnyyEPL2009}
{ V.~B. Zabolotnyy, A.~A. Kordyuk, D.~S. Inosov, D.~V. Evtushinsky,
  \textit{et~al}., \href{http://stacks.iop.org/0295-5075/86/47005}{
\newblock EPL {\bf 86}, 47005 (6pp) (2009)}.}

\bibitem{KordyukPRB2005}
{ A.~A. Kordyuk, S.~V. Borisenko, A.~Koitzsch, J.~Fink, \textit{et~al}.,
  \href{http://link.aps.org/abstract/PRB/v71/e214513}{
\newblock Phys. Rev. B {\bf 71}, 214513 (2005)}.}

\bibitem{PashitskiiLTP2008}
{ E.~A. Pashitskii and V.~I. Pentegov,
  \href{http://link.aip.org/link/?LTP/34/113/1}{
\newblock Low Temperature Physics {\bf 34}, 113 (2008)}.}

\end{thebibliography}

\end{document}